\pdfoutput=1
\documentclass[a4paper]{llncs} 
\usepackage{tex/mystyle}

\urldef{\mailsa}\path|{linus3,watanabe}@is.titech.ac.jp|
\urldef{\mailsb}\path|frejohk@chalmers.se|
\newcommand{\keywords}[1]{\par\addvspace\baselineskip
\noindent\keywordname\enspace\ignorespaces#1}

\begin{document}

\mainmatter  

\title{Generalized Shortest Path Kernel on Graphs}

\titlerunning{Generalized Shortest Path Kernel}

\author{Linus Hermansson\inst{1} 
\and Fredrik D. Johansson\inst{2} \and Osamu Watanabe\inst{1} }


\institute{Tokyo Institute of Technology, Tokyo, Japan,
\and Chalmers University of Technology, Gothenburg, Sweden \\
\mailsa \\ 
\mailsb }

\maketitle
\setcounter{footnote}{0}

\begin{abstract}
We consider
the problem of classifying graphs using graph kernels.
We define a new graph kernel,
called the generalized shortest path kernel,
based on the number and length of shortest paths between nodes.
For our example classification problem,
we consider the task of classifying random graphs
from two well-known families,
by the number of clusters they contain.
We verify empirically that
the generalized shortest path kernel
outperforms the original shortest path kernel on a number of datasets.
We give a theoretical analysis for explaining our experimental results.
In particular,
we estimate distributions of
the expected feature vectors for the shortest path kernel
and the generalized shortest path kernel,
and we show some evidence
explaining why our graph kernel outperforms the shortest path kernel
for our graph classification problem.

\end{abstract}

\section{Introduction}
\label{sec:Introduction}

Classifying graphs into different classes depending on their structure is a problem that has been studied for a long time and that has many useful applications~\cite{bilgin2007cell,borgwardt2005protein,kong2010semi,kudo2004application}. By classifying graphs researchers have been able to solve important problems such as to accurately predict the toxicity of chemical compounds~\cite{kudo2004application}, classify if human tissue contains cancer or not~\cite{bilgin2007cell}, predict if a particular protein is an enzyme or not~\cite{borgwardt2005protein}, and many more. 

It is generally regarded that the number of self-loop-avoiding paths between all pairs of nodes of a given graph is useful for understanding the structure of the graph~\cite{havlin1982theoretical,liskiewicz2003complexity}.
 Computing the number of such paths between all nodes is however a computationally hard task (usually \#P-hard). Counting only the number of shortest paths between node pairs is however possible in polynomial time and such paths at least avoid cycles, which is why some researchers have considered shortest paths a reasonable substitute. When using standard algorithms to compute the shortest paths between node pairs in a graph we also get, as a by-product, the \emph{number} of such shortest paths between all node pairs. Taking this number of shortest paths into account when analyzing the properties of a graph could provide useful information and is what our approach is built upon.

One popular technique for classifying graphs is by using a \emph{support vector machine} (SVM) classifier with graph kernels. This approach has proven successful for classifying several types of graphs~\cite{borgwardt2005shortest,borgwardt2005protein,hermansson2013entity}. Different graph kernels can however give vastly different results depending on the types of graphs that are being classified. Because of this it is useful to analyze which graph kernels that works well on which types of graphs. Such an analysis contributes to understanding when graph kernels are useful and on which types of graphs one can expect a good result using this approach. 
In order to classify graphs, graph kernels that consider many different properties have been proposed. Such as graph kernels considering all walks~\cite{gartner2003graph}, shortest paths~\cite{borgwardt2005shortest}, small subgraphs~\cite{shervashidze2009efficient}, global graph properties~\cite{johansson2014global}, and many more. Analyzing how these graph kernels perform for particular datasets, gives us the possibility of choosing graph kernels appropriate for the particular types of graphs that we are trying to classify. 

One particular type of graphs, that appears in many applications, are graphs with a cluster structure. Such graphs appear for instance when considering graphs representing social networks.
In this paper, in order to test how well our approach works,
we test its performance on the problem of classifying graphs
by the number of clusters that they contain.
More specifically,
we consider two types of models for generating random graphs,
the Erd\H{o}s-R\'enyi model~\cite{bollobas1998random}
and the planted partition model~\cite{kollaspectra},
where we use the Erd\H{o}s-R\'enyi model to generate graphs with one cluster
and the planted partition model
to generate graphs with two clusters
(explained in detail in Sect. \ref{sec:gmodel}).
The example task considered in this paper
is to classify whether a given random graph is generated
by the Erd\H{o}s-R\'enyi model or by the planted partition model.

For this classification problem,
we use the standard SVM
and compare experimentally
the performance of the SVM classifier,
with the \emph{shortest path} (SP) kernel, and with our new \emph{generalized shortest path} (GSP) kernel.
In the experiments we generate datasets with 100 graphs generated according to the Erd\H{o}s-R\'enyi model and 100 graphs generated according to the planted partition model. Different datasets use different parameters for the two models. The task is then, for any given dataset, to classify graphs as coming from the Erd\H{o}s-R\'enyi model or the planted partition model, where we consider the supervised machine learning setting with 10-fold cross validation.
We show that the SVM classifier that uses our GSP kernel outperforms the SVM classifier that uses the SP kernel, on several datasets.

Next we give
some theoretical analysis of the random feature vectors
of the SP kernel and the GSP kernel, for the random graph models used in our experiments.
We give an approximate estimation
of expected feature vectors for the SP kernel
and show that the expected feature vectors are relatively close
between graphs with one cluster and graphs with two clusters.
We then analyze the distributions of component values
of expected feature vectors for the GSP kernel,
and we show some evidence that
the expected feature vectors have a different structure
between graphs with one cluster and graphs with two clusters.

The remainder of this paper is organized as follows. In Sect. \ref{sec:prelim} we introduce notions and notations that are used throughout the paper. Section \ref{sec:kernels} defines already existing and new graph kernels. In Sect. \ref{sec:gmodel} we describe the random graph models that we use to generate our datasets. Section \ref{sec:Experiments} contains information about our experiments and experimental results. In Sect. \ref{sec:Analysis} we give an analysis explaining why our GSP kernel outperforms the SP kernel on the used datasets. Section \ref{sec:conclusions} contains our conclusions and suggestions for future work.

\section{Preliminaries}
\label{sec:prelim}

Here we introduce necessary notions and notation for our technical discussion.
Throughout this paper
we use symbols $G$, $V$, $E$  (with a subscript or a superscript)
to denote graphs, sets of nodes, and sets of edges respectively.
We fix $n$ and $m$
to denote the number of nodes and edges of considered graphs.
By $|S|$ we mean the number of elements of the set $S$.

We are interested in the length and number of shortest paths.
In relation to the kernels we use for classifying graphs,
we use {\em feature vectors} for expressing such information.
For any graph $G$,
for any $d\ge1$,
let $n_d$ denote
the number of pairs of nodes of $G$ with a shortest path of length $d$
(in other words, distance $d$ nodes).
Then we call
a vector $\vecvsp=[n_1,n_2,\ldots]$
a {\em SPI feature vector}.
On the other hand,
for any $d,x\ge1$,
we use $n_{d,x}$ to denote
the number of pairs of nodes of $G$ that have $x$ number of shortest paths of length $d$,
and we call
a vector $\vecvgsp=[n_{1,1},n_{1,2},\ldots , n_{2,1} \ldots]$ a {\em GSPI feature vector}.
Note that
$n_d=\sum_x n_{d,x}$.
Thus,
a GSPI feature vector is a more detailed version of a SPI feature vector.
In order to simplify our discussion
we often use feature vectors by considering shortest paths
from any fixed node of $G$.
We will clarify which version we use in each context. By $\myE[ \vecvsp ]$ and $\myE[ \vecvgsp ]$ we mean the expected SPI feature vector and the expected GSPI feature vector, for some specified random distribution. Note that the expected feature vectors are equal to 
$[\myE[n_d]]_{d\ge1}$ and $[\myE[n_{d,x} ] ]_{d\ge1,x\ge1}$.

It should be noted that
the SPI and the GSPI feature vectors are computable efficiently.
For example,
we can use
Dijkstra's algorithm \cite{dijkstra1959note}
for each node in a given graph,
which gives all node pairs' shortest path length
(i.e. a SPI feature vector)
in time $\mathcal{O}(nm + n^2\log n)$.
Note that by using Dijkstra's algorithm to compute the shortest path from a fixed source node to any other node, the algorithm actually needs to compute {\em all} shortest paths between the two nodes, to verify that it really has found a shortest path. In many applications, however, we are only interested in obtaining one shortest path for each node pair, meaning that we do not store all other shortest paths for that node pair. It is however possible to store the number of shortest paths between all node pairs, {\bf without increasing the running time of the algorithm}, 
meaning that we can compute the GSPI feature vector in the same time as the SPI feature vector. Note that for practical applications, it might be wise to use a binning scheme for the number of shortest paths, where we consider numbers of shortest paths as equal if they are close enough. For example instead of considering the numbers of shortest paths $\{1 , 2 ... 100 \}$, as different. We could consider the intervals $\{[1,10], [11,20] ... [91,100] \}$ as different and consider all the numbers inside a particular interval as equal. Doing this will reduce the dimension of the GSPI feature vector, which could be useful since the number of shortest paths in a graph might be large for dense graphs.

We note that the graph kernels used in this paper can be represented explicitly as inner products of finite dimensional feature vectors. We choose to still refer to them as \emph{kernels}, because of their relation to other graph kernels.

\section{Shortest Path Kernel and Generalized Shortest Path Kernel}
\label{sec:kernels}

A graph kernel is a function $k(G_1,G_2)$ on pairs of graphs, which can be represented as an inner product $k(G_1,G_2) = \langle \phi(G_1), \phi(G_2) \rangle_{\mathcal{H}}$ for some mapping $\phi(G)$ to a Hilbert space $\mathcal{H}$, of possibly infinite dimension. In many cases, graph kernels can be thought of as similarity functions on graphs.
Graph kernels have been used as tools
for using SVM classifiers for graph classification problems~\cite{borgwardt2005shortest,borgwardt2005protein,hermansson2013entity}.

The kernel
that we build upon in this paper is
the \emph{shortest path} (SP) kernel,
which compares graphs based
on the shortest path length of all pairs of nodes~\cite{borgwardt2005shortest}.
By $D(G)$ we denote
the multi set of shortest distances between all node pairs in the graph $G$.
For two given graphs $G_1$ and $G_2$,
the SP kernel is then defined as:
\begin{equation*}
K_{\rm SP}(G_{1}, G_{2})
=\sum_{d_1\in D(G_{1})}\;\sum_{d_2\in D(G_{2})}k(d_1,d_2),
\end{equation*}
where $k$ is a positive definite kernel~\cite{borgwardt2005shortest}.
One of the most common kernels for $k$ is the indicator function,
as used in Borgwardt and Kriegel~\cite{borgwardt2005shortest}.
This kernel compares shortest distances for equality.
Using this choice of $k$ we obtain the following definition of the SP kernel:
\begin{equation}
\label{eq:SPKernelInd}
K_{\rm SPI}(G_{1},G_{2})
=\sum_{d_1\in D(G_{1})}\;\sum_{d_2\in D(G_{2})}
\mathds{1} \left[d_1=d_2\right].
\end{equation}
We call this version of the SP kernel
the \emph{shortest path index} (SPI) kernel.
It is easy to check that
$K_{\rm SPI}(G_{1},G_{2})$ is simply
the inner product of the SPI feature vectors of $G_1$ and $G_2$.

We now introduce our new kernel,
the \emph{generalized shortest path} (GSP) kernel,
which is defined by using {\em also} the number of shortest paths.
For a given graph $G$,
by $ND(G)$ we denote
the multi set of numbers of shortest paths between all node pairs of $G$.
Then the GSP kernel is defined as:
\begin{equation*}
K_{\rm GSP}(G_{1}, G_{2})
=
\sum_{d_1\in D(G_1)}\;\sum_{d_2\in D(G_2)}\;
\sum_{t_1\in ND(G_1)}\;\sum_{t_2\in ND(G_2)}
k(d_1,d_2,t_1,t_2),
\end{equation*}  
where $k$ is a positive definite kernel.
A natural choice for $k$ would be again
a kernel where we consider node pairs as equal
if they have the same shortest distance \emph{and}
the same number of shortest paths.
Resulting in the following definition,
which we call the \emph{generalized shortest path index} (GSPI) kernel.
\begin{equation}
\label{eq:GSPKernelInd}
K_{\rm GSPI}(G_{1}, G_{2}) =
\sum_{d_1\in D(G_1)}\;\sum_{d_2\in D(G_2)}\;
\sum_{t_1\in ND(G_1)}\;\sum_{t_2\in ND(G_2)}
\mathds{1}\left[d_1=d_2\right]\mathds{1}\left[t_1=t_2\right]
\end{equation}
It is easy to see
that this is equivalent to 
the inner product of the GSPI feature vectors of $G_1$ and $G_2$.

\section{Random Graph Models}
\label{sec:gmodel}

We investigate
the advantage of our GSPI kernel over the SPI kernel
for a synthetic random graph classification problem.
Our target problem is
to distinguish random graphs having two relatively ``dense parts'',
from simple graphs generated by the Erd\H{o}s-R\'enyi model.
Here by ``dense part''
we mean a subgraph that has more edges
in its inside compared with its outside.

For any edge density parameter $p$, $0<p<1$,
the  Erd\H{o}s-R\'enyi model (with parameter $p$) denoted by $G(n,p)$
is to generate a graph $G$ (of $n$ nodes)
by putting an edge between each pair of nodes
with probability $p$ independently at random.
On the other hand,
for any $p$ and $q$, $0<q<p<1$,
the {\em planted partition model}~\cite{kollaspectra},
denoted by $G(n/2,n/2,p,q)$
is to generate a graph $G=(V^+\cup V^-,E)$ (with $|V^+|=|V^-|=n/2$)
by putting an edge between each pair of nodes $u$ and $v$
again independently at random
with probability $p$ if both $u$ and $v$ are in $V^+$ (or in $V^-$)
and with probability $q$ if $u\in V^+$ and $v\in V^-$
(or, $u\in V^-$ and $v\in V^+$).

Throughout this paper,
we use the symbol $p_1$ to denote the edge density parameter
for the Erd\H{o}s-R\'enyi model
and $p_2$ and $q_2$ to denote the edge density parameters
for the planted partition model.
We want to have $q_2<p_2$
while keeping the expected number of edges the same
for both random graph models
(so that
one cannot distinguish random graphs by just couting the number of edges).
It is easy to check
that this requirement is satisfied by setting
%
%
\begin{equation}
\label{eq:pandq}
p_2=(1+\alpha_0)p_1,
{\rm~~and~~}
q_2=2p_1-p_2-2(p_1 - p_2)/n
\end{equation}
for some constant $\alpha_0$, $0<\alpha_0<1$.
We consider the ``sparse'' situation for our experiments and analysis,
and assume that $p_1=c_0/n$ for sufficiently large constant $c_0$.  
Note that
we may expect with high probability, that when $c_0$ is large enough, a random graph generated by both models have a large connected component but might not be fully connected~\cite{bollobas1998random}.
In the rest of the paper,
a random graph generated by $G(n,p_1)$ is called
a {\em one-cluster graph}
and a random graph generated by $G(n/2,n/2,p_2,q_2)$ is called
a {\em two-cluster graph}.

For a random graph, the
SPI/GSPI feature vectors are random vectors.
For each $z\in\{0,1\}$,
we use $\vecvspz$ and $\vecvgspz$ to denote
random SPI and GSPI feature vectors of a $z$-cluster graph.
We use $\numz{d}$ and $\numz{d,x}$
to denote respectively
the $d$th and $(d,x)$th component of $\vecvspz$ and $\vecvgspz$.
For our experiments and analysis,
we consider their expectations $\vecvspEz$ and $\vecvgspEz$,
that is,
$[\numEz{d}]_{d\ge1}$ and $[\numEz{d,x}]_{d\ge1,x\ge1}$.
Note that
$\myE[\numz{d,x}]$ is the expected {\em number of node pairs}
that have $x$ number of shortest paths of length $d$;
not to be confused
with the expected number of distance $d$ shortest paths.

\section{Experiments}
\label{sec:Experiments}

In this section we compare the performance of the GSPI kernel with the SPI kernel on datasets where the goal is to classify if a graph is a one-cluster graph or a two-cluster graph.

\subsection{Generating Datasets and Experimental Setup}
\label{sec:generation}

All datasets are generated using the models $G(n,p_1)$ and $G(n/2,n/2,p_2,q_2)$, described above. We generate 100 graphs from the two different classes in each dataset. 
$q_2$ is chosen in such a way that the expected number of edges is the same for both classes of graphs.
 Note that when $p_2 = p_1$, the two-cluster graphs actually become one-cluster graphs where all node pairs are connected with the same probability, meaning that the two classes are indistinguishable. The bigger difference there is between $p_1$ and $p_2$, the more different the one-cluster graphs are compared to the two-cluster graphs.
In our experiments we generate graphs where $n \in \{200,400,600,800,1000\}$, $np_1=c_0 = 40$ and $p_2 \in \{ 1.2p_1, 1.3p_1, 1.4p_1, 1.5p_1\} $. Hence $p_1 = 0.2$ for $n=200$, $p_1=0.1$ for $n=400$ etc.

In all experiments we calculate the normalized feature vectors for all graphs. By normalized we mean that each feature vector $\vecvsp$ and $\vecvgsp$ is normalized by its Euclidean norm. This means that the inner product between two feature vectors always is in $[0,1]$.
We then train an SVM using 10-fold cross validation and evaluate the accuracy of the kernels. We use Pegasos~\cite{shalev2011pegasos} for solving the SVM.

\subsection{Results}

Table \ref{table:accuracy} shows the accuracy of both kernels, using 10-fold cross validation, on the different datasets. As can be seen neither of the kernels perform very well on the datasets where $p_2=1.2 p_1$. This is because the two-cluster graphs generated in this dataset are almost the same as the one-cluster graphs. As $p_2$ increases compared to $p_1$, the task of classifying the graphs becomes easier. As can be seen in the table the GSPI kernel outperforms the SPI kernel on nearly all datasets. In particular, on datasets where $p_2 = 1.4 p_1$, the GSPI kernel has an increase in accuracy of over 20\% on several datasets. When $n=200$ the increase in accuracy is over 40\%!
Although the shown results are only for datasets where $c_0=40$, experiments using other values for $c_0$ gave similar results. 

\begin{figure}[tb]
\centering
    \includegraphics[width=1.0\columnwidth, keepaspectratio=true]{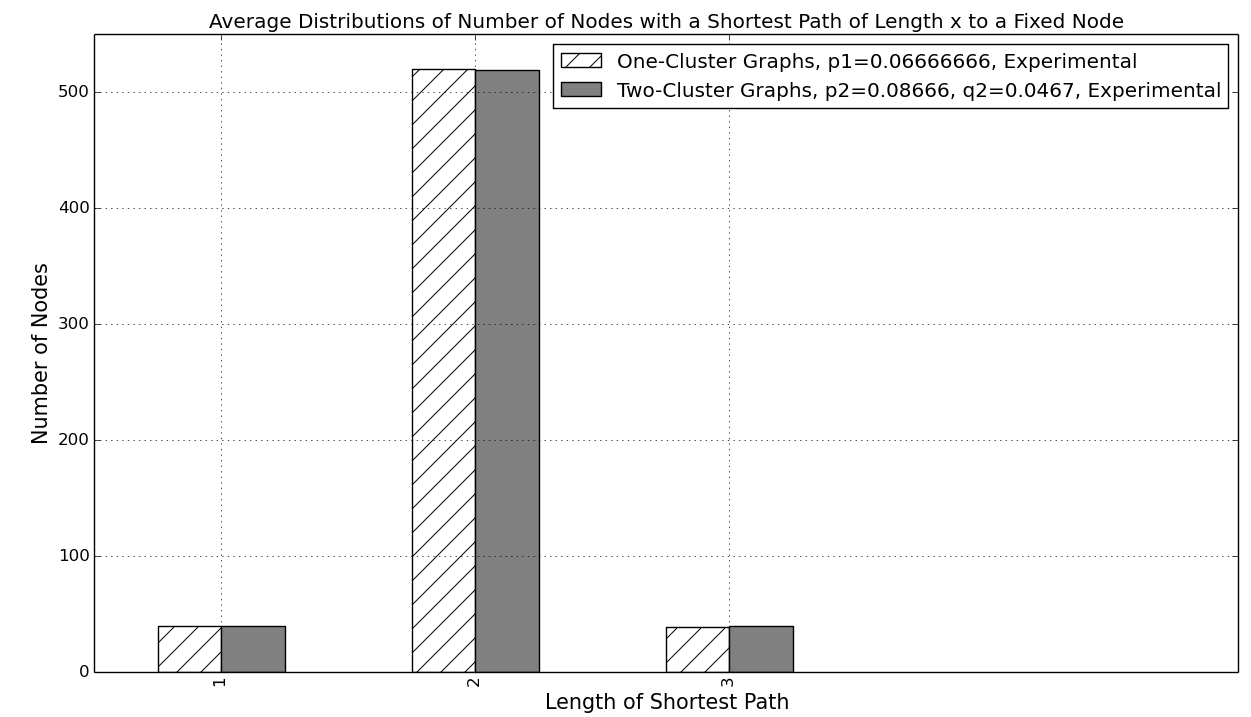}
\caption{%
Average distributions of number of nodes with a shortest path of length $x$ to a fixed node. The distributions have been averaged for each node in the graph and also averaged over 100 randomly generated graphs, for both classes of graphs. The graphs used the parameters $n=600$, $p_1=0.06667$ for one-cluster graphs and $p_2=0.08667$, $q_2=0.04673$ for two-cluster graphs.
}
\label{fig:len_sp}
\end{figure}

\begin{figure}[tb]
\centering
    \includegraphics[width=1.0\columnwidth, keepaspectratio=true]{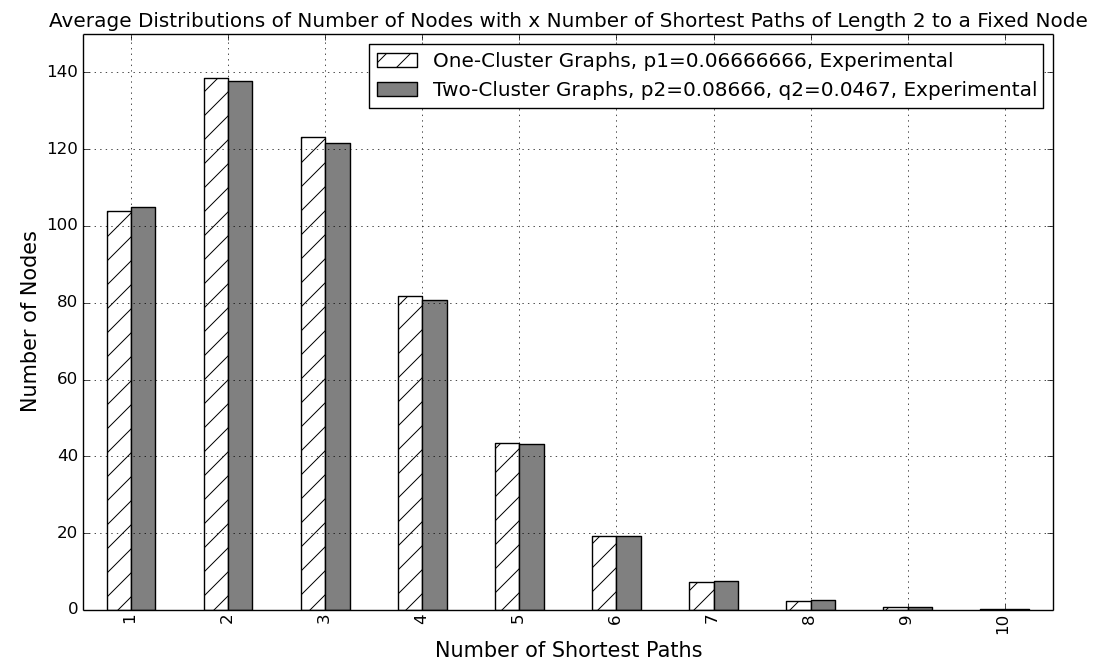}
\caption{%
Average distributions of number of nodes with $x$ number of shortest paths of length 2 to a fixed node. The distributions have been averaged for each node in the graph and also averaged over 100 randomly generated graphs, for both classes of graphs. The graphs used to generate this figure are the same as in Fig. \ref{fig:len_sp}.
}
\label{fig:num_sps}
\end{figure}

One reason that our GSPI kernel is able to classify graphs correctly when the SPI kernel is not, is because the feature vectors of the GSPI kernel, for the two classes, are a lot more different than for the SPI kernel. In Fig. \ref{fig:len_sp} we have plotted the SPI feature vectors, for a fixed node, for both classes of graphs and one particular dataset. 
By feature vectors for a fixed node we mean that the feature vectors contains information for one fixed node, instead of node pairs, so that for example, $n_d$ from $\vecvsp=[n_1,n_2,\ldots]$,  contains the number of nodes that are at distance $d$ from one fixed node, instead of the number of node pairs that are at distance $d$ from each other. The feature vector displayed in Fig. \ref{fig:len_sp} is the average feature vector, for any fixed node, and averaged over the 100 randomly generated graphs of each type in the dataset. The dataset showed in the figure is when the graphs were generated with $n=600$, the one-cluster graphs used $p_1=0.06667$, the two-cluster graphs used $p_2=0.08667$ and $q_2=0.04673$, this corresponds to, in Table \ref{table:accuracy}, the dataset where $n=600$, $p_2=1.3p_1$, this dataset had an accuracy of $60.5\%$ for the SPI kernel and $67.0\%$ for the GSPI kernel. 
As can be seen in the figure there is almost no difference at all between the average SPI feature vectors for the two different cases. 
In Fig. \ref{fig:num_sps}  we have plotted the subvectors $\numtwoone{x}_{x\ge1}$ of $\vecvgspfst$ and  $\numtwotwo{x}_{x\ge1}$ of $\vecvgspsnd$, for a fixed node, for the same dataset as in Fig. \ref{fig:len_sp}. 
The vectors contain the number of nodes at distance 2 from the fixed node with $x$ number of shortest paths, for one-cluster graphs and two-cluster graphs respectively. The vectors have been averaged for each node in the graph and also averaged over the 100 randomly generated graphs, for both classes of graphs, in the dataset. 
As can be seen the distributions of such numbers of nodes are at least distinguishable for several values of $x$, when comparing the two types of graphs. 
This motivates why the SVM is able to distinguish the two classes better using the GSPI feature vectors than the SPI feature vectors.

\begin{table}[tbp]
\caption{The accuracy of the SPI kernel and the GSPI kernel using 10-fold cross validation. The datasets where $p_2 = 1.2 p_1$ are the hardest and the datasets where $p_2 = 1.5 p_1$ are the easiest. Very big increases in accuracy are marked in bold.}
\centering
\begin{tabular}{|lccc|}
\hline
Kernel & \multicolumn{1}{l}{$n$} & \multicolumn{1}{l}{$p_2$} & \multicolumn{1}{l|}{Accuracy}\\ \hline
SPI            & 200        & $\{ 1.2p_1, 1.3p_1, 1.4p_1, 1.5p_1 \}$       & $\{ 52.5\%  , 55.5\% ,  {\bf 54.5\% }  {\bf 56.5\%} \} $     \\ 
GSPI         & 200        & $\{ 1.2p_1, 1.3p_1, 1.4p_1, 1.5p_1 \}$       & $ \{   52.5 \% , 64.0 \% , {\bf 99.0 \%} , {\bf 100.0 \% } \} $ \\ 

\hline

SPI           & 400   & $\{ 1.2p_1, 1.3p_1, 1.4p_1, 1.5p_1 \}$       &   $\{  55.5\% ,  63.5\%  , {\bf 75.5\% } , 95.5\% \} $  \\ 
GSPI           & 400        & $\{ 1.2p_1, 1.3p_1, 1.4p_1, 1.5p_1 \}$       & $ \{ 54.0\%  ,  62.0 \% , {\bf 96.5 \% } , 100.0 \% \} $ \\ 

\hline

SPI           & 600        & $\{ 1.2p_1, 1.3p_1, 1.4p_1, 1.5p_1 \}$       & $\{  58.0\% , 60.5\%, {\bf 75.5\%} , 93.5\%\}$ \\ 
GSPI          & 600        & $\{ 1.2p_1, 1.3p_1, 1.4p_1, 1.5p_1 \}$       & $ \{  58.0 \% , 67.0\% , {\bf 94.0 \% } , 100.0 \% \}  $ \\ 

\hline

SPI       & 800        & $\{ 1.2p_1, 1.3p_1, 1.4p_1, 1.5p_1 \}$       & $\{ 57.5 \%, 59.0\%  , 72.0\%, 98.0 \%  \} $ \\ 
GSPI          & 800        & $\{ 1.2p_1, 1.3p_1, 1.4p_1, 1.5p_1 \}$       & $\{ 57.5\% ,  58.0 \%  , 82.0 \% ,  100.0 \% \} $ \\ 

\hline

SPI       & 1000        & $\{ 1.2p_1, 1.3p_1, 1.4p_1, 1.5p_1 \}$       & $ \{  53.5 \% , 55.0 \% ,{\bf  66.0 \% }, 98.5 \%  \}  $    \\ 
GSPI          & 1000        & $\{ 1.2p_1, 1.3p_1, 1.4p_1, 1.5p_1 \}$       & $\{ 55.0\% , 62.0\% , {\bf 87.5\% }, 100.0\%  \} $  \\ 
\hline
\end{tabular}

\label{table:accuracy}
\end{table}

\def\OMIT#1{}

\section{Analysis}
\label{sec:Analysis}

In this section
we give some approximated analysis of random feature vectors
in order to give theoretical support for our experimental observations.
We first show that one-cluster and two-cluster graphs
have quite similar SPI feature vectors (as their expectations).
Then we next show some evidence
that there is a non-negligible difference in their GSPI feature vectors.
Throughout this section,
we consider feature vectors
defined by considering only paths from any fixed source node $s$.
Thus,
for example,
$\numfst{d}$ is the number of nodes at distance $d$ from $s$
in a one-cluster graph,
and $\numsnd{d,x}$ is the number of nodes
that have $x$ shortest paths of length $d$ to $s$ in a two-cluster graph.
\OMIT{
For the following analysis,
the difference between ``walk'' and ``path'' becomes crucial.
Recall that
a {\em walk} is simply a sequence of neighboring nodes,
while a {\em path} also has the requirement
that no node in the path may appear more than once.}

Here we introduce a way to state an approximation.
For any functions $a$ and $b$ depending on $n$,
we write
$a\myapprox b$ by which we mean
\[
b\left(1-{c\over n}\right)
<a<b\left(1+{c\over n}\right)
\]
holds for some constant $c>0$ and sufficiently large $n$.
We say that
$a$ and $b$ are {\em relatively $\myrelclose$-close}
if $a\myapprox b$ holds.
Note that
this closeness notion
is closed under constant number of additions/subtractions and multiplications.
For example,
if $a\myapprox b$ holds,
then we also have $a^k\myapprox b^k$ for any $k\ge1$
that can be regarded as a constant w.r.t.\ $n$. 
In the following
we will often use this approximation.

\subsection{Approximate Comparison of SPI Feature Vectors}

We consider relatively small\footnote{%
This smallness assumption is for our analysis,
and we believe that the situation is more or less the same for any $d$.}
distances $d$
so that $d$ can be considered as a small constant w.r.t.\ $n$.
We show that
$\numEfst{d}$ and $\numEsnd{d}$ are similar in the following sense.

\begin{theorem}
\label{similartheorem}
For any constant $d$,
we have
$\numEfst{d} \in \numEsnd{d}(1 \pm \frac{2}{c_0 - 1}) $, holds within our $\myapprox$ approximation when $c_0 \geq 2 + \sqrt{3}$.
%
\end{theorem}

\noindent
{\bf Remark.}~
For deriving this relation
we assume a certain independence
on the existence of two paths in $G$;
see the argument below for the detail. Note that this difference between $\numEfst{d}$ and $\numEsnd{d}$ vanishes for large values of $c_0$.
\bigskip

\begin{proof}
First consider a one-cluster graph $G=(V,E)$ and analyze $\numEfst{d}$.
For this analysis,
consider any target node $t$ ($\ne s$) of $G$ (we consider this target node to be a fixed node to begin with),
and estimate first the probability $\Ffst_d$
that there exists at least one path of length $d$ from $s$ to $t$.
Let $A_{u,v}$ be the event that an edge $\{u,v\}$ exists in $G$,
and let $\myW$ denote
the set of
all paths (from $s$ to $t$)
expressed
by a permutation $(v_1,\ldots,v_{d-1})$ of nodes in $V\setminus\{s,t\}$.
For each tuple $(v_1,\ldots,v_{d-1})$ of $\myW$,
the event $A_{s,v_1}\land A_{v_1,v_2}\land\ldots\land A_{v_{d-1},t}$
is called
the event that
the path (from $s$ to $t$)
specified by $(s,v_1,\ldots,v_{d-1},t)$ exists in $G$
(or, more simply,
{\em
the existence of one specific path}).
Then the probability $\Ffst_d$ is expressed by
\begin{equation}
\label{eq:Ffst}
\Ffst_d
=\myP\left[\,
\bigvee_{(v_1,\ldots,v_{d-1})\in \myW}
A_{s,v_1}\land A_{v_1,v_2}\land\ldots\land A_{v_{d-1},t}\,\right]
\end{equation}
Clearly,
the probability of the existence of one specific path is $p_1^d$,
and the above probability can be calculated
by using the inclusion-exclusion principle.
Here we follow the analysis of Fronczak et al.~\cite{fronczak2004average}
and assume that
every specific path exists independently. Note that the number of dependent paths can be big when the length of a path is long, therefore this assumption is only reasonable for short distances $d$. 
To simplify the analysis\footnote{%
Clearly,
this is a rough approximation;
nevertheless,
it is enough for asymptotic analysis w.r.t.\ the $\myrelclose$-closeness.
For smaller $n$,
we may use a better approximation from \cite{fronczak2004average},
which will be explained in Subsection~\ref{sec:inex}.}
we only consider the first term of the inclusion-exclusion principle.
That is,
we approximate $\Ffst_d$ by
\begin{eqnarray}
\Ffst_d
&\approx&
\label{eq:Ffst_approxA}
\sum_{(v_1,\ldots,v_{d-1})\in \myW}
\Pr\left[\,
A_{s,v_1}\land A_{v_1,v_2}\land\ldots\land A_{v_{d-1},t}\,\right]\\
&=&
\nonumber
|\myW|p_1^d
~=~
(n-2)(n-3)\cdots(n-(2+d-2))p_1^d
~\myapprox~
n^{d-1}p_1^d,
\end{eqnarray}
where the last approximation relation holds since $d$ is constant.
From this approximation,
we can approximate
the probability $\ffst_d$ that
$t$ has a shortest path of length $d$ to $s$.
For any $d\ge1$,
let $\EventFfst_d$ be the event that there exists at least one path of length $d$, {\em or less}, between $s$ and $t$, and let $\Eventffst_d$ be
the event 
that there exists a shortest path of length $d$
between $s$ and $t$. Then 
\begin{equation}
F_d \leq \myP[\EventFfst_d] \leq \displaystyle\sum_{i=1}^{d} \Ffst_i.
\end{equation}
Note that 
\begin{eqnarray}
\displaystyle\sum_{i=1}^{d} \Ffst_i  ~ &\myapprox& \displaystyle\sum_{i=1}^{d} n^{i-1} p_1^i 
= n^{d-1} p_1^d ( \displaystyle\sum_{i=0}^{d-1} \frac{1}{(np_1)^i} ) \nonumber \\
&\leq & n^{d-1}p_1^{d} (\frac{n p_1}{np_1 - 1}) 
= n^{d-1}p_1^d (1 + \frac{1}{np_1 -1}). \nonumber
\end{eqnarray}
Since $np_1= c_0 \geq 1$, it follows that 
\begin{equation*}
n^{d-1}p_1^d (1 + \frac{1}{np_1 -1}) = n^{d-1}p_1^d (1 + \frac{1}{c_0 -1}).
\end{equation*}
While 
$F_d \myapprox n^{d-1}p_1^{d}$. Thus we have within our $\myapprox$ approximation, that 
\begin{equation*}
n^{d-1}p_1^{d} \leq \myP[\EventFfst_d] \leq n^{d-1}p_1^d (1 + \frac{1}{c_0 -1}).
\end{equation*}

It is obvious that $\ffst_d=\Pr[\Eventffst_d]$, 
note also that
$\EventFfst_d$ $=$ $\Eventffst_d\lor \EventFfst_{d-1}$
and that the two events $\Eventffst_d$ and $\EventFfst_{d-1}$ are disjoint.
Thus,
we have
$\Pr[\EventFfst_d]$ $=$ $\Pr[\Eventffst_d]$+$\Pr[\EventFfst_{d-1}]$,
which is equivalent to
\[
 n^{d-1} p_1^{d} - n^{d-2}p_1^{d-1}(1 + \frac{1}{c_0 -1} ) \leq \ffst_d \leq n^{d-1}p_1^d(1 + \frac{1}{c_0 -1} ) - n^{d-2} p_1^{d-1}.
\]
Since $\ffst_d$ is the probability that
there is a shortest path of length $d$ from $s$ to any fixed $t$,
it follows that $\numEfst{d}$,
i.e.,
the expected number of nodes that have a shortest path of length $d$ to $s$,
can be estimated by
\begin{equation*}
\label{eq:numEfst}
\numEfst{d}
=
(n-1)\ffst_d
\myapprox
n\ffst_d.
\end{equation*}
Which gives that
\begin{equation}
\label{complicated}
 n^{d} p_1^{d} - n^{d-1}p_1^{d-1}(1 + \frac{1}{c_0 -1} ) \leq \numEfst{d} \leq n^{d}p_1^d(1 + \frac{1}{c_0 -1} ) - n^{d-1} p_1^{d-1}.
\end{equation}
holds within our $\myapprox$, approximation. 

We may rewrite the above equation using the following equalities 
\begin{align}
 n^{d} p_1^{d} - n^{d-1}p_1^{d-1}(1 + \frac{1}{c_0 -1} ) &= n^d p_1^d - n^{d-1} p_1^{d-1} - \frac{n^{d-1}p_1^{d-1}}{c_0 - 1} \nonumber \\
&=  (n^d p_1^d - n^{d-1} p_1^{d-1}) ( 1 - \frac{ \frac{ n^{d-1}p_1^{d-1} }{c_0 - 1}} {  n^d p_1^d - n^{d-1} p_1^{d-1} } )  \nonumber \\
&=   (n^d p_1^d - n^{d-1} p_1^{d-1}) ( 1 - \frac{ 1 } { (c_0 - 1)^2} ), \label{simple1}
\end{align}
and 
\begin{align}
n^{d}p_1^d(1 + \frac{1}{c_0 -1} ) - n^{d-1} p_1^{d-1} &= n^{d} p_1^{d} - n^{d-1} p_1^{d-1} + \frac{n^{d} p_1^{d}}{c_0 - 1} \nonumber \\
&= (n^{d} p_1^{d} - n^{d-1} p_1^{d-1})( 1 + \frac{ \frac{n^{d} p_1^{d}}{c_0 - 1} } { n^{d} p_1^{d} - n^{d-1} p_1^{d-1}} ) \nonumber \\
&= (n^{d} p_1^{d} - n^{d-1} p_1^{d-1})( 1 + \frac{c_0}{(c_0 - 1)^2} ). \label{simple2}
\end{align}
Substituting  (\ref{simple1}) and (\ref{simple2}) into (\ref{complicated}) we get
\begin{equation}
\label{bounds1}
\lowerb \leq \numEfst{d} \leq \upperb .
\end{equation}
We will later use these bounds to derive the theorem.

We now analyze a two-cluster graph and $\numEsnd{d}$.
Let us assume first that $s$ is in $V^+$.
Again we fix a target node $t$ to begin with.
Here we need to consider
the case that the target node $t$ is also in $V^+$ and
the case that it is in $V^-$.
Let $\Fsndp_d$ and $\Fsndm_d$  be 
the probabilities that
$t$ has at least one path of length $d$ to $s$ in the two cases.
Then for the first case,
the path starts from $s \in V^+$ and ends in $t \in V^+$,
meaning that the number of times that
the path crossed from one cluster to another
(either from $V^+$ to $V^-$ or $V^-$ to $V^+$) has to be even.
Thus the probability of one specific path existing
is $p_2^{d-k}q_2^k$ for some even $k$, $0\le k\le d$.
Thus,
the first term of the inclusion-exclusion principle
(the sum of the probabilities of all possible paths) then becomes
\begin{equation*}
\Fsndp_d
\myapprox
\left(n\over2\right)^{d-1}
\sum_{{\rm even}~k=0}^{d}{d\choose k}p_2^{d-k}q_2^k,
\end{equation*}
where the number of paths is approximated as before,
i.e.,
$|V^+\setminus\{s,t\}|\cdot(|V^+\setminus\{s,t\}|-1)
\cdots((|V^+\setminus\{s,t\}|-(d-2))$ is approximated by $(n/2)^{d-1}$.
We can similarly analyze the case where $t$ is in $V^-$ to obtain
\begin{equation*}
\Fsndm_d
\myapprox
\left(n\over2\right)^{d-1}
\sum_{{\rm odd}~k=1}^{d}{d\choose k}p_2^{d-k}q_2^k.
\end{equation*}
Since both cases ($t \in V^+$, or $t \in V^-$) are equally likely, the average probability of there being a path of length $d$, between $s$ and $t$, in a two-cluster graph is
\begin{eqnarray*}
\frac{\Fsndp_d+\Fsndm_d}{2}
&\myapprox&
\left(n\over2\right)^{d-1}
\sum_{{\rm even}~k=0}^{d}
{d\choose k}\frac{p_2^{d-k}q_2^k}{2}
+
\left(n\over2\right)^{d-1}
\sum_{{\rm odd}~k=1}^{d}{d\choose k}\frac{p_2^{d-k}q_2^k}{2} \\
&=&
\left(n\over2\right)^{d-1}
\sum_{k=0}^{d}
{d\choose k}\frac{p_2^{d-k}q_2^k}{2}
=
\left(n\over2\right)^{d-1}\frac{(p_2+q_2)^d}{2}.
\end{eqnarray*}
Note here that
$p_2+q_2$ $\myapprox$ $2p_1$ from our choice of $q_2$ (see (\ref{eq:pandq})).
Thus, we have
\begin{equation*}
\frac{\Fsndp_d+\Fsndm_d}{2}
\myapprox 
\left(n\over2\right)^{d-1}\frac{(2p_1)^d}{2} = n^{d-1} p_1^d.
\end{equation*}
Which is exactly the same as in the one cluster case, see (\ref{eq:Ffst_approxA}). Thus we have
\begin{equation}
\label{bounds2}
\lowerb \leq \numEsnd{d} \leq \upperb.
\end{equation}

Using this we now prove the main statement of the theorem, namely that 
\begin{equation*}
\numEfst{d} \in \numEsnd{d} ( 1 \pm \frac{2}{c_0 - 1}).
\end{equation*}
To prove the theorem we need to prove the following two things
\begin{align}
\numEfst{d} &\leq \numEsnd{d} ( 1 + \frac{ 2} {c_0-1}), \text{ and} \label{prove1} \\
\numEfst{d} &\geq \numEsnd{d} ( 1 - \frac{ 2} {c_0-1}) \label{prove2}
\end{align}
The proof of (\ref{prove1}) can be done by using (\ref{bounds1}) and (\ref{bounds2}).
\begin{align}
&\numEfst{d} \leq \numEsnd{d} + \frac{ 2 \numEsnd{d} } {c_0-1}  \nonumber \\
&\Leftrightarrow \upperb \leq \lowerb   \nonumber \\
&~ ~ ~ ~ + \frac{ 2 \lowerb}{c_0 - 1}  \nonumber \\
&\Leftrightarrow 1 + \frac{c_0}{(c_0 - 1)^2} \leq 1 - \frac{1}{(c_0-1)^2} + \frac{2}{c_0 - 1} - \frac{2}{(c_0-1)^3} \nonumber \\ 
&\Leftrightarrow \frac{c_0 + 1}{c_0-1} + \frac{2}{(c_0 - 1)^2} \leq 2.
\end{align}
Which holds when $c_0 \geq 2 + \sqrt{3} \approx 3.7$. The proof of (\ref{prove2}) is similar and shown below.
\begin{align}
&\numEfst{d} \geq \numEsnd{d} - \frac{ 2 \numEsnd{d} } {c_0-1} \nonumber \\
&\Leftrightarrow \lowerb \geq \upperb  \nonumber \\
& ~ ~ ~ ~ - \frac{2 \lowerb}{c_0 - 1} \nonumber \\
&\Leftrightarrow 1 - \frac{1}{(c_0 - 1)^2} \geq 1 + \frac{c_0}{(c_0 -1)^2} - \frac{2}{c_0 -1} + \frac{2}{ (c_0 -1)^3} \nonumber \\
&\Leftrightarrow 2 \geq \frac{c_0+ 1}{c_0 - 1} + \frac{2}{(c_0 -1)^2}.
\end{align}
Which again holds when $c_0 \geq 2 + \sqrt{3} \approx 3.7$. This completes the proof of the theorem.
\noindent\hfill
$\square$
\end{proof}

\subsection{Heuristic Comparison of GSPI Feature Vectors}

We compare in this section
the expected GSPI feature vectors $\vecvgspEfst$ and $\vecvgspEsnd$,
that is,
$[\numEfst{d,x}]_{d\ge1,x\ge1}$ and $[\numEsnd{d,x}]_{d\ge1,x\ge1}$,
and show evidence
that they have some non-negligible difference.
Here we focus on the distance $d=2$ part of the GSPI feature vectors,
i.e.,
subvectors $[\numtwoEz{x}]_{x\ge1}$ for $z\in\{1,2\}$.
Since it is not so easy to analyze
the distribution of the values $\numtwoEz{1},\numtwoEz{2},\ldots$,
we introduce some ``heuristic'' analysis.

We begin with a one-cluster graph $G$,
and let $V_2$ denote the set of nodes of $G$
with distance 2 from the source node $s$.
Consider any $t$ in $V_2$,
and for any $x\ge1$,
we estimate the probability that
it has $x$ number of shortest paths of length 2 to $s$.
Let $V_1$ be the set of nodes at distance 1 from $s$.
Recall that
$G$ has $(n-1)\ffst_1 \myapprox np_1$ nodes in $V_1$ on average,
and we assume that
$t$ has an edge from some node in $V_1$
each of which corresponds to a shotest path of distance 2 from $s$ to $t$.
We now assume for our ``heuristic'' analysis that
$|V_1|=np_1$ and that
an edge between each of these distance 1 nodes and $t$ exists
with probability $p_1$ independently at random.
Then $x$ follows the binomial distribution $\myBin(np_1,p_1)$,
where by $\myBin(N,p)$
we mean a random number of heads that we have
when flipping a coin that gives heads with probability $p$
independently $N$ times.
Then for each $x\ge1$,
$\numtwoEfst{x}$,
the expected number of nodes of $V_2$
that have $x$ shortest paths of length 2 to $s$,
is estimated by
\[
\numtwoEfst{x}
\approx
\sum_{t\in V_2}\myP\bigl[\,\myBin(np_1,p_1)=x\,\bigr]
=
\numEfst{2}
\cdot
\myP\bigl[\,\myBin(np_1,p_1)=x\,\bigr],
\]
by assuming
that $|V_2|$ takes its expected value $\numEfst{2}$.
Clearly
the distribution of values of vector $[\numtwoEfst{x}]_{x\ge1}$
is proportional to $\myBin(np_1,p_1)$,
and it has one peak at $\xpeakfst=np_1^2$,
since the mean of a binomial distribution, $\myBin(N,p)$ is $Np$.

Consider now a two-cluster graph $G$.
We assume that our start node $s$ is in $V^+$.
For $d\in\{1,2\}$,
let $V_d^+$ and $V_d^-$ denote respectively
the set of nodes in $V^+$ and $V^-$ with distance $d$ from $s$.
Let $V_2=V_2^+\cup V_2^-$.
Again we assume that
$V_1^+$ and $V_1^-$ have respectively $np_2/2$ and $nq_2/2$ nodes
and that
the numbers of edges
from $V_1^+$ and $V_1^-$ to a node in $V_2$ follow binomial distributions.
Note that we need to consider two cases here,
$t \in V_2^+$ and $t \in V_2^-$.
First consider the case that the target node $t$ is in $V_2^+$.
In this case
there are two types of shortest paths.
The first type of paths goes from $s$ to $V^+_1$ and then to $t \in V_2^+$.
The second type of shortest path goes from $s$ to $V^-_1$ and then to $t \in V_2^+$.
Based on this we get
\begin{eqnarray*}
\ptwosndp{x}
&:=&
\myP\bigl[\,\mbox{$t$ has $x$ shortest paths}\,\bigr]\\
&=&
\myP\left[\,\myBin\left({n\over2}p_2,p_2\right)
+\myBin\left({n\over2}q_2,q_2\right)=x\,\right]\\
&\approx&
\myP\left[\,\myN\left({n\over2}p_2^2,\sigma_1^2\right)
+\myN\left({n\over2}q_2^2,\sigma_2^2\right)\in[x-0.5,x+0.5] \,\right]\\
&=&
\myP\left[\,\myN
\left({n(p_2^2+q_2^2)\over2},\sigma_1^2+\sigma_2^2\right)
\in[x-0.5,x+0.5] \, \right],
\end{eqnarray*}
where we use the
normal distribution $\myN(\mu,\sigma^2)$
to approximate each binomial distribution
so that we can express their sum by a normal distribution
(here we omit specifying $\sigma_1$ and $\sigma_2$).
For the second case where $t\in V^-_2$,
with a similar argument,
we derive
\[
\ptwosndm{x}
:=
\myP\bigl[\,\mbox{$t$ has $x$ shortest paths}\,\bigr]
=
\myP\left[\,\myN
\left(np_2q_2,\sigma_3^2 + \sigma_4^2 \right)\in[x-0.5,x+0.5] \,  \right].
\]

Note that the first case ($t \in V_2^{+}$), happens with probability $|V_2^{+}| / (|V_2^{+}| + |V_2^{-}|)$. The second case  ($t \in V_2^{-}$), happens with probability $|V_2^{-}| / (|V_2^{+}| + |V_2^{-}|)$.
Then again we may approximate
the $x$th component of the expected feature subvector
by 
\begin{equation*}
  \numtwoEsnd{x} \approx \frac{\myE[ |V_2^{+}|] }{ \myE[|V_2^{+}|] + \myE[|V_2^{-}|]} \myE[|V_2^+|]\ptwosndp{x}+  \frac{\myE[ |V_2^{-}|] }{ \myE[|V_2^{+}|] + \myE[|V_2^{-}|]}    \myE[|V_2^-|]\ptwosndm{x}.
\end{equation*}

We have now arrived at the key point in our analysis.
Note that
the distribution of values of vector $[\numtwoEsnd{x}]_{x\ge1}$ follows
the mixture of two distributions,
namely,
$\myN(n(p_2^2+q_2^2)/2,\sigma_1^2+\sigma_2^2)$
and $\myN(np_2q_2,\sigma_3^2 + \sigma_4^2)$,
with weights $\myE[ |V_2^{+}|] / (\myE[|V_2^{+}|] + \myE[|V_2^{-}|])$ and $\myE[ |V_2^{-}|] / (\myE[|V_2^{+}|] + \myE[|V_2^{-}|])$.
Now we estimate
the distance
between the two peaks $\xpeaksndp$ and $\xpeaksndm$ of these two distributions.
Note that
the mean of a normal distribution $\myN(\mu,\sigma^2)$ is simply $\mu$.
Then we have
\begin{eqnarray*}
\xpeaksndp-\xpeaksndm
&=&
\frac{n}{2}(p_2^2+q_2^2)-np_2q_2
=
\frac{n}{2}(p_2-q_2)^2\\
&\myapprox&
\frac{n}{2}(p_2-(2p_1-p_2))^2
=
\frac{n}{2}(2p_2 - 2p_1)^2 \\
&\myapprox& 
2n(p_1(1 + \alpha_0)-p_1)^2=2n(p_1\alpha_0)^2
=
2n\alpha_0^2p_1^2
\end{eqnarray*}
Note that
$q_2 \myapprox 2p_1 - p_2$ holds (from (\ref{eq:pandq}));
hence,
we have
$p_1$ $\myapprox$ $(p_2+q_2)/2$ $\ge$ $\sqrt{p_2q_2}$,
and we approximately have
$p_1^2$ $\ge$ $p_2q_2$.
By using this,
we can bound the difference between these peaks by
\begin{equation*}
\xpeaksndp-\xpeaksndm
\myapprox
2n\alpha_0^2p_1^2
\ge
2\alpha_0^2\xpeaksndm.
\end{equation*}
That is,
these peaks have non-negligible relative difference.

From our heuristic analysis
we may conclude that the
two vectors $[\numtwoEfst{x}]_{x\ge1}$ and $[\numtwoEsnd{x}]_{x\ge1}$
have different distributions of their component values.
In particular,
while the former vector has only one peak,
the latter vector has a double peak shape (for large enough $\alpha_0)$.
Note that this difference does not vanish even when
$c_0$ is big. This means that the GSPI feature vectors are different for one-cluster
graphs and two-clusters graphs, even when $c_0$ is big, which is not the case for the
SPI feature vectors, since their difference vanishes when $c_0$ is big. This provides
evidence as to why our GSPI kernel performs better than the SPI kernel

Though this is a heuristic analysis,
we can show some examples that
our observation is not so different from experimental results.
In Fig.~\ref{fig:exp_approx}
we have plotted both our approximated vector of $[\numtwoEsnd{x}]_{x\ge1}$
(actually we have plotted
the mixed normal distribution that gives this vector)
and the corresponding experimental vector obtained
by generating graphs according to our random model.
In this figure the double peak shape can clearly be observed,
which provides empirical evidence supporting our analysis.
This experimental vector
is the average vector for each fixed source node in random graphs,
which is averaged over 500 randomly generated graphs
with the parameters $n=400$, $p_2=0.18$, and $q_2=0.0204$.
(For these parameters,
we need to use a better approximation of (\ref{eq:Ffst}) explained
in the next subsection to derive the normal distribution of this figure.)

\subsection{Inclusion-Exclusion Principle}
\label{sec:inex}
\begin{figure}[tb]
\centering
\includegraphics[width=1.0\columnwidth, keepaspectratio=true]{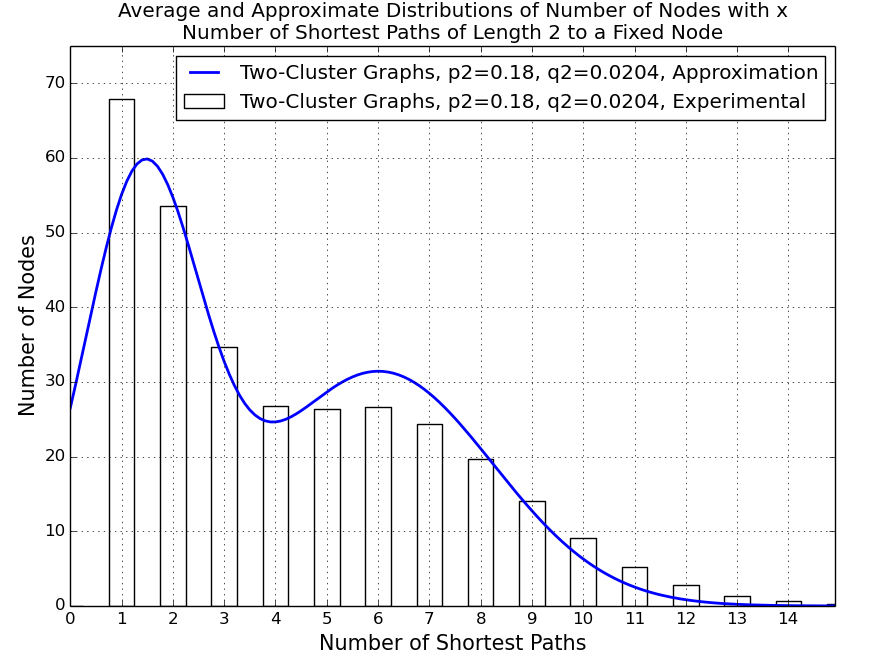}
\caption{%
Average experimental and approximate distributions of number of nodes
with $x$ number of shortest paths of length 2 from a fixed node.
The experimental distribution has been averaged
for each node in the graph and also averaged
over 500 randomly generated graphs.
Graphs used had parameters $n=400$, $p_2=0.18$ and $q_2=0.0204$.
(For computing the graph of the approximate distribution,
we used a more precise approximation for $\Ffst_d$
because our $n$ is not large enough, see Sect. \ref{sec:inex} for details.)}
\label{fig:exp_approx}
\end{figure}

Throughout the analysis,
we have always used the first term of
the inclusion-exclusion principle to estimate (\ref{eq:Ffst}).
Although this works well for expressing our analytical result,
where we consider the case where $n$ is big.
When applying the approximation for graphs with a small number of nodes,
it might be necessary to consider a better approximation of
the inclusion-exclusion principle.
For example,
we in fact used
the approximation from \cite{fronczak2004average}
for deriving the mixed normal distributions of Fig.~\ref{fig:exp_approx}.
Here for completeness,
we state this approximation as a lemma
and give its proof that is outlined in \cite{fronczak2004average}.

\begin{lemma}
\label{lem:bound}
Let $E_1,E_2,\ldots,E_l$ be mutually independent events
such that $\myP[E_i]\leq\epsilon$ holds for all $i$, $1\le i\le l$.
Then we have
\begin{equation}
\label{eq:lemfron}
\myP\left[\,\bigcup^{l}_{i=1}E_i\,\right]
=
1-\myexp\left(-\displaystyle\sum^l_{i=1}\myP[E_i]\right)-Q.
\end{equation}
Where
\begin{equation}
\label{eq:errorQ}
-\displaystyle\sum_{k=0}^{l+1} \frac{(l \epsilon)^k}{k!} + (1+\epsilon)^{l} 
\le Q \le
\displaystyle\sum_{k=0}^{l+1} \frac{(l \epsilon)^k}{k!} - (1+\epsilon)^{l} .
\end{equation}
\end{lemma}

\noindent
{\bf Remark.}~
The above bound for the error term $Q$ is
slightly weaker than the one in \cite{fronczak2004average},
but it is sufficient enough for many situations,
in particular for our usage.
In our analysis of (\ref{eq:Ffst})
each $E_i$ corresponds to an event that one specific path exists.
Recall that we assumed that all paths exists independently.
\bigskip

\begin{proof}
Using the definition of the inclusion-exclusion principle we get
\begin{equation}
\label{eq:bigcup}
\myP\left[\,\bigcup^{l}_{i=1}E_i\,\right]
=
\sum^{l}_{k=1} (-1)^{k+1} S(k),
\end{equation}
where each $S(k)$ is defined by
\begin{equation*}
S(k)
=
\sum_{1 \leq i_1 < \ldots < i_k \leq l}
\myP[E_{i_1}] \myP[E_{i_2}] \cdots \myP[E_{i_k}]
=
\sum_{1 \leq i_1 < \ldots < i_k \leq l}
P_{i_1} P_{i_2}\cdots P_{i_k}.
\end{equation*}
Here and in the following
we denote each probability $\myP[E_i]$ simply by $P_i$.

First we show that
\begin{equation}
\label{eq:Sk}
S(k)
=
{1\over k!}\left(\sum_{i=1}^{l}P_i\right)^k-Q_k,
\end{equation}
where
\begin{equation}
\label{eq:Qk}
0\le Q_k\le\left(\frac{l^k}{k!}-\binom{l}{k}\right)\epsilon^k.
\end{equation}
To see this
we introduce
two index sequence sets $\Gamma_k$ and $\Pi_k$ defined by
\begin{eqnarray*}
\Gamma_k
&=&
\{\,(i_1,\ldots,i_k)\,:\,
\mbox{$i_j\in\{1,\ldots,l\}$ for all $j$, $1\le j\le k$}\,\},\\
\Pi_k
&=&
\{\,(i_1,\ldots,i_k)\in\Gamma_k\,:\,
\mbox{$i_j\ne i_{j'}$ for all $j,j'$, $1\le j<j'\le k$}\,\}.
\end{eqnarray*}
Then it is easy to see that
\[
\left(\sum_{i=1}^{l}P_i\right)^k
=
\sum_{(i_1,\ldots,i_k)\in\Gamma_k}P_{i_1}\cdots P_{i_k},
{\rm~~and~~}
k!S(k)
=
\sum_{(i_1,\ldots,i_k)\in\Pi_k}P_{i_1}\cdots P_{i_k}.
\]
Thus,
we have
\begin{eqnarray*}
k!Q_k
=
\left(\sum_{i=1}^{l}P_i\right)^k-k!S(k)
&=&
\sum_{(i_1,\ldots,i_k)\in\Gamma_k\setminus\Pi_k}P_{i_1}\cdots P_{i_k}\\
&\le&
|\Gamma_k\setminus\Pi_k|\epsilon^k
=
\bigl(l^k-l(l-1)\cdots(l-k+1)\bigr)\epsilon^k,
\end{eqnarray*}
which gives bound (\ref{eq:Qk}) for $Q_k$.

Now from (\ref{eq:bigcup}) and (\ref{eq:Sk})
we have
\begin{equation*}
1-\myP\left[\,\bigcup^{l}_{i=1}E_i\,\right]
=
\sum^{l}_{k=0}{(-1)^k\over k!}\left(\sum_{i=1}^{l}P_i\right)^k
+\sum^{l}_{k=1}(-1)^{k+1}Q_k.
\end{equation*}
Here we note that
the sum $\sum^{l}_{k=0}(-1)^k/k!(\sum_{i=1}^{l}P_i)^k$ is
the first $l+1$ terms of
the MacLaurin expansion of $\myexp(-\sum_{i=1}^lP_i)$.
Hence,
the error term $Q$ of (\ref{eq:lemfron}) becomes
\begin{equation*}
Q
=
- \sum_{k\ge l+1}{(-1)^k\over k!}\left(\sum_{i=1}^{l}P_i\right)^k
+ \sum^{l}_{k=1}(-1)^{k+1}Q_k.
\end{equation*}
We now derive an upper bound for $Q$.
\begin{eqnarray}
Q
&\le&
- \sum_{k\ge l+1}\frac{(-1)^{k}}{k!}\left(\sum_{i=1}^{l}P_i\right)^k
+\sum^{l}_{k=1}Q_k \nonumber \\
&\le &
\frac{(l \epsilon)^{l+1}}{(l+1)!}
+ \displaystyle\sum_{k=1}^{l} \frac{(l \epsilon)^k}{k!} - \displaystyle\sum_{k=1}^{l} \binom{l}{k} \epsilon^k \nonumber \\
&=&\displaystyle\sum_{k=0}^{l+1} \frac{(l \epsilon)^k}{k!} - (1+\epsilon)^{l}. \nonumber
\end{eqnarray}
This proves the upper bound on $Q$, the proof for the lower bound of $Q$ is completely analogous. Thus, the lemma holds.
\noindent
$\square$
\end{proof}

\section{Conclusions and Future Work}
\label{sec:conclusions}
 
We have defined a new graph kernel, based on the number of shortest paths between node pairs in a graph. The feature vectors of the GSP kernel do not take longer time to calculate than the feature vectors of the SP kernel. The reason for this is the fact that the number of shortest paths between node pairs is a by-product of using Dijkstra's algorithm to get the length of the shortest paths between all node pairs in a graph. The number of shortest paths between node pairs {\bf does} contain relevant information for certain types of graphs. In particular we showed in our experiments that the GSP kernel, which also uses the number of shortest paths between node pairs, outperformed the SP kernel, which only uses the length of the shortest paths between node pairs, at the task of classifying graphs as containing one or two clusters. We also gave an analysis motivating why the GSP kernel is able to correctly classify the two types of graphs when the SP kernel is not able to do so. 

Future research could examine the distribution of the random feature vectors $\vecvsp$ and $\vecvgsp$, for random graphs, that are generated using the planted partition model, and have \emph{more} than two clusters. Although we have only given experimental results and an analysis for graphs that have either one or two clusters, preliminary experiments show that the GSPI kernel outperforms the SPI kernel on such tasks as e.g. classifying if a random graph contains one or four cluster, if a random graph contains two or four clusters etc. It would be interesting to see which guarantees it is possible to get, in terms of guaranteeing that the $\vecvgsp$ vectors are different and the $\vecvsp$ vectors are similar, when the numbers of clusters are not just one or two.

\bibliographystyle{abbrv}

\end{document}